\documentclass[floatfix,twocolumn,showpacs,preprintnumbers,amsmath,amssymb,
pra,superscriptaddress,longbibliography]{revtex4-2}

\usepackage{tabularray}
\UseTblrLibrary{booktabs} 
\usepackage{xtab}
\usepackage{color}
\usepackage[usenames,dvipsnames,svgnames,table]{xcolor}
\usepackage[colorlinks=true,linkcolor=blue,urlcolor=blue,citecolor=blue]{hyperref}
\usepackage{mathtools}
\usepackage{graphicx}
\usepackage{dcolumn}
\usepackage{lipsum}
\usepackage{bm}
\usepackage{subfigure}
\usepackage{amssymb}
\usepackage{multirow}
\usepackage{amsmath}
\usepackage{braket}
\usepackage{csquotes}
\graphicspath{{plots/}}
 \usepackage{lipsum}
\usepackage{mathrsfs}
\usepackage{siunitx}
\usepackage{fixdif}
\usepackage{listings}

\usepackage{enumitem,amssymb}
\newlist{todolist}{itemize}{2}
\setlist[todolist]{label=$\square$}
\usepackage{pifont}
\newcommand{\beq}{\begin{equation}}
\newcommand{\eeq}{\end{equation}}
\newcommand{\bea}{\begin{eqnarray}}
\newcommand{\eea}{\end{eqnarray}}

\newcommand{\kv}{{\bf k}}

\begin{document}
\title{
X-Ray Diagnostics Analysis Verification and Exploration (xDAVE) Code for the Prediction and Interpretation of X-Ray Thomson Scattering Experiments
}

\author{Hannah M.~Bellenbaum}
\email{h.bellenbaum@hzdr.de}
\affiliation{Center for Advanced Systems Understanding (CASUS), Helmholtz-Zentrum Dresden-Rossendorf (HZDR), D-02826 G\"orlitz, Germany}
\affiliation{Institut f\"ur Physik, Universit\"at Rostock, D-18057 Rostock, Germany}

\author{David~A.~Chapman}
\affiliation{First Light Fusion Ltd., Oxford OX5 1QU, UK}

\author{Maximilian~P.~B\"ohme}
\affiliation{Lawrence Livermore National Laboratory (LLNL), California 94550 Livermore, USA}

\author{Thomas~Gawne}
\affiliation{Center for Advanced Systems Understanding (CASUS), Helmholtz-Zentrum Dresden-Rossendorf (HZDR), D-02826 G\"orlitz, Germany}

\author{Sebastian~Schwalbe}
\affiliation{Institute of Radiation Physics, Helmholtz-Zentrum Dresden-Rossendorf (HZDR), D-01328 Dresden, Germany}

\author{Willow~M.~Martin}
\affiliation{SLAC National Accelerator Laboratory, Stanford University, California  94025 Menlo Park, USA}
\affiliation{Physics Department, Stanford University, Stanford, California 94305, USA}

\author{Michael Bussmann}
\affiliation{Center for Advanced Systems Understanding (CASUS), Helmholtz-Zentrum Dresden-Rossendorf (HZDR), D-02826 G\"orlitz, Germany}

\author{Dirk~O.~Gericke}
\affiliation{Centre for Fusion, Space and Astrophysics, University of Warwick, Coventry CV4 7AL, UK}

\author{Uwe~Hernandez~Acosta}
\affiliation{Center for Advanced Systems Understanding (CASUS), Helmholtz-Zentrum Dresden-Rossendorf (HZDR), D-02826 G\"orlitz, Germany}


\author{Jan Vorberger}
\affiliation{Institute of Radiation Physics, Helmholtz-Zentrum Dresden-Rossendorf (HZDR), D-01328 Dresden, Germany}

\author{Tobias Dornheim}
\affiliation{Institute of Radiation Physics, Helmholtz-Zentrum Dresden-Rossendorf (HZDR), D-01328 Dresden, Germany}
\affiliation{Center for Advanced Systems Understanding (CASUS), Helmholtz-Zentrum Dresden-Rossendorf (HZDR), D-02826 G\"orlitz, Germany}

\begin{abstract}
X-ray Thomson scattering (XRTS) is a common diagnostic used in the warm dense matter (WDM) regime to estimate plasma parameters like density, temperature and charge state.
Experimental analysis typically relies on a forward model to obtain estimates for these parameters, as the measured spectrum is a convolution of the dynamic structure factor (DSF) and the source-instrument function.
The Chihara decomposition, where the spectrum is separated into contributions from bound and free electrons, is commonly used to estimate DSFs in the WDM regime, as it allows for the fast calculation of DSFs and therefore can easily be applied in a large-scale parameter optimization.
Due to the limited availability of XRTS codes, in this work we present the ``\textbf{X}-ray \textbf{D}iagnostics, \textbf{A}nalysis, \textbf{V}erification and \textbf{E}xploration`` (\texttt{xDAVE}) code, designed to quickly estimate DSFs using the Chihara decomposition and analyse experimental spectra.
The code is validated by re-analysing an experiment with isochorically heated beryllium at the OMEGA Laser Facility.
In addition, we demonstrate the applicability of the code to plan experiments and predict scattering spectra through the coupling to a ray-tracing code.
Lastly, the importance of accounting for the energy-dependence of spectrometer instrument functions is demonstrated by comparing ray-tracing simulations  to the standard convolution for strongly compressed Beryllium shots at the National Ignition Facility similar to previously published results.
\end{abstract}
\maketitle

\section{Introduction}

Warm dense matter (WDM) describes an extreme state of matter, commonly characterised as the transition phase between condensed matter and plasma physics~\cite{wdm_roadmap_preprint}.
It occurs ubiquitously in the universe, in many astrophysical objects such as planetary interiors and stars, and plays a vital role in further advancements in material sciences.
For example, the interior of Jupiter has been extensively studied (e.g.~\cite{Militzer_2008,Militzer_2016}), but producing an accurate model heavily relies on the theoretical understanding of the different hydrogen-helium mixtures under WDM conditions.
Recent experiments have further highlighted the applicability of the WDM field in the manufacturing and understanding of novel materials~\cite{wdm_roadmap_preprint}.
For instance, Kraus et al.~\cite{Kraus2016,Kraus2017} have demonstrated the creation of nanodiamonds from shocked graphite and WDM experiments have been used to qualify crystal structures in materials like hydrocarbons and metallic oxides~\cite{Conway_2021}.
Perhaps the most important application of WDM however lies within inertial fusion energy (IFE).
During the implosion process, a typical inertial confinement (ICF) target traverses the WDM regime on its way to ignition~\cite{hu_ICF}.
Therefore, extensive work has focussed on accurately characterising materials relevant for future inertial fusion energy targets under these conditions by experimentally diagnosing conditions both in-flight (see e.g.~\cite{Kritcher_2011}) and laser-heated (see e.g.~\cite{Fletcher_2014}).
Particularly potential ablator materials like beryllium or diamond that commonly make up the implosion capsule~\cite{Kritcher_2022} are studied to validate theoretical models and improve upon them.
Further developing experimental diagnostics is therefore vital in the development of IFE as a viable energy source and to further validate simulations, for example radiation-hydrodynamics~\cite{wdm_roadmap_preprint,Marinak_POP_2024}.

X-ray Thomson scattering (XRTS) has emerged as one of the most capable diagnostics in the WDM regime to characterise thermodynamic conditions and microphsysics of plasmas at a number of facilities~\cite{siegfried_review}.
These include x-ray free electron lasers (XFELs) like the European XFEL ~\cite{Voigt_POP_2021,bespalov2026experimentalevidencebreakdownuniformelectrongas} or the Linac Coherent Light Source (LCLS) at the SLAC National Accelerator Laboratory ~\cite{kraus_xrts,Sperling_PRL_2015,Martin_2025}, and laser facilities such as the National Ignition Facility (NIF) ~\cite{Kritcher2020,Tilo_Nature_2023} and the OMEGA Laser at the Laboratory for Laser Energetics (LLE) ~\cite{Boehly_1997,Ross_2006,Kritcher_2011,Fletcher_2013,Poole_2024-06}.
XRTS probes the electronic dynamic structure factor (DSF) of a system, which contains information, for example, about electronic correlations and the density response~\cite{kremp2005quantum,Dornheim_review}.
The DSF also contains all relevant thermodynamic properties of a system, so XRTS as a diagnostic is in principle capable of simultaneously measuring electron and ion temperature, mass density and charge state~\cite{siegfried_review}.
It has also been used to estimate the impact of continuum lowering on the binding energies of carbon~\cite{kraus_xrts} and electron-ion equilibration rates~\cite{Fletcher_Frontiers_2022}, making it a very versatile diagnostic method.
At the NIF, XRTS has also been utilised in characterizing the implosion process of beryllium capsules that are typically used in ICF targets and validate radiation-hydrodynamics simulations~\cite{Tilo_Nature_2023}, giving insights into the extreme states of matter reached during the implosion stage.

While XRTS has been proven to be an important and multipurpose diagnostic approach, it is complicated by the fact that the DSF is not directly measured.
Instead, recorded spectra are, to very good approximation, a convolution of the DSF and the source-instrument function (SIF) taking into account effects due to the crystal, the detector and the source spectrum~\cite{Gawne_2024-09}.
Noise and experimental uncertainties make a direct deconvolution often impractical and difficult~\cite{Voigt_POP_2021}.
Experimental analysis of XRTS spectra therefore commonly relies on a forward-model to obtain best-fit parameters through an optimization scan or Markov chain Monte Carlo (MCMC) sampling~\cite{Kasim_POP_2019}.
The choice of forward-model usually depends on computational feasibility and accuracy.
~\emph{Ab initio} simulation tools like density functional theory (DFT) or path integral Monte Carlo (PIMC), while being the most accurate tools to model the WDM regime~\cite{new_POP}, are computationally very expensive and often limited to comparatively small particle numbers.
Both DFT and PIMC have recently been used to extract density estimates from a NIF capsule implosion by comparing against the elastic scattering signal~\cite{Dornheim_2025_POP,Dornheim_2025_NatCommun}.
Time-dependent DFT (TD-DFT) has also been applied to model compressed beryllium by Baczewski~\emph{et al.}~\cite{dynamic2}, to re-analyse NIF beryllium spectra by Sch\"orner~\emph{et al.}~\cite{Schoerner_PRE_2023} and to interpret plasmon measurements in shocked aluminium by Bespalov~\emph{et al.}~\cite{bespalov2026experimentalevidencebreakdownuniformelectrongas}.
High computational costs, however, mean they are generally unsuitable for multi-dimensional MCMC sampling that is necessary to extract temperature, density and charge state estimates from an XRTS spectrum~\cite{Kasim_POP_2019,Poole_POP_2022}.
Frequently, the forward model applied in the analysis of experimental datasets therefore makes use of the Chihara decomposition, which applies a chemical picture to the scattering processes where free and bound electrons are assumed to be clearly separable~\cite{Chihara_1987,siegfried_review}.
This allows for quick calculations of the DSF using simple, often analytic models that are very accessible and computationally cheap compared to \emph{ab initio} simulation tools, making large parameter scans feasible~\cite{siegfried_review}.

The availability of open-source XRTS codes for the experimental analysis is currently limited.
To address this, we present the ``X-ray Diagnostics, Analysis, Verification and Exploration`` (\texttt{xDAVE}) code to calculate dynamic structure factors within the chemical picture.
The aim of this code is to provide a modular approach to the components within the Chihara decomposition, and make the code applicable not only to experimental analysis but also to compare against~\emph{ab initio} simulation tools to determine shortcomings of these chemical models and improve them.
In addition, this new code was designed to work in combination with model-free analysis techniques recently published by Dornheim~\emph{et al.} (see e.g.~\cite{Dornheim_T_2022,Dornheim_T_follow_up}).
Their work has focused on interpreting results in the imaginary-time domain, where the deconvolution of an experimentally spectrum is straightforward, offering a model-free pathway to the temperature through the detailed balance relation~\cite{Dornheim_T_2022}.
\texttt{xDAVE} gives direct access to the imaginary-time correlation function (ITCF) required for this analysis, which further allows the tool to be used in comparison against PIMC data, for example to extend previous work on estimating charge state and ionization potential depression of warm dense hydrogen~\cite{Bellenbaum_2025_itcf} to higher-Z materials and DFT.

To further improve on the experimental analysis, we demonstrate the coupling between \texttt{xDAVE} and a ray-tracing code to fully capture the asymmetry and photon energy-dependence of the instrument function (IF) component of the SIF.
Both the model-free temperature analysis and the forward-fitting rely on an accurate description of the SIF, particularly in the upshifted regime where the asymmetry of the IF with respect to photon energy can affect estimated temperatures~\cite{Gawne_2024-09}.
To remove uncertainties due to the shape of the SIF, ray-tracing the DSFs can be useful in both the planning of future experiments and the analysis as part of the forward model.

This paper will initially focus on the theoretical description of the DSF in the reduced chemical picture (see Section~\ref{sec:theory}), before describing the new XRTS code and the coupling to a ray-tracing code in Section~\ref{sec:method}.
We then validate the code by re-analysing a beryllium XRTS spectrum obtained at the OMEGA Laser Facility~\cite{Doeppner_2009} (see Section~\ref{subsec:be-omega}).
We then apply the combined ray-tracing DSF approach to first demonstrate predictive capabilities for a typical European XFEL experiment in Section~\ref{subsec:eu-xfel} before demonstrating the importance of this approach on the analysis of beryllium capsule implosions at the NIF(see Section~\ref{subsec:nif-implosions}).
We will conclude by summarising our results and giving  outlooks on future code development and applications.

\section{Theory}~\label{sec:theory}

The measured XRTS signal can, to a good approximation~\cite{Gawne_preprint_2026}, be expressed as a convolution of the DSF with a combined source-instrument function as
~\cite{siegfried_review}
\begin{equation}
    P(\kv,\omega) \propto R(\omega) \ast S_{ee}(\kv,\omega) \,,
\end{equation}
thus determining plasma conditions like the temperature, density and mean charge state relies on assumptions in modelling both the DSF and the SIF.

The DSF in the spectral representation is defined as~\cite{hansen2013theory}
\begin{equation}
    S_{ee}(\kv,\omega) = \frac{1}{2 \pi} \int_{-\infty}^{\infty} dt F(\kv, t) \exp{(i\omega t)}\,,
\end{equation}
where the intermediate scattering function in turn is defined as
\begin{equation}
    F(\kv,t) = \frac{1}{N} \langle \varrho(\kv,t) \varrho(-\kv,0) \rangle \,,
\end{equation}
in terms of the Fourier component of the microscopic density $\varrho(\kv,t)$.
The wavevector $\kv$ is the momentum transfer in the system, defined by the difference between the incoming and the scattered particle momenta $\kv = \kv_i - \kv_s$.
The energy transfer $\omega$ is similarly defined as $\omega=\omega_i - \omega_s$.
Note that frequency and energy are often treated interchangeably. Here, we define $\omega$ as the energy transfer, it is related to the frequency $w_\text{freq}$ as $w_\text{freq} = w / \hbar$.

Using the fluctuation-dissipation theorem~\cite{Kubo_1966}
\begin{equation}
    S_{ee}(\kv,\omega) = -\frac{1}{\pi n_e} \frac{1}{1 - \exp{(-\beta \omega)}} \mbox{Im} [\chi_{ee}(\kv,\omega)] \,,
\end{equation}
for the inverse electron temperature $\beta=1/k_b T_e$, we relate the DSF with the density response function $\chi_{ee}(\kv,\omega)$
\begin{equation}
    \chi_{ee}(\kv, \omega) 
    = -\frac{i}{\hbar}\,\Theta(t)\, \left\langle \left[ \varrho(\mathbf{k}, t), \varrho(-\mathbf{k}, 0) \right] \right\rangle \,.
\end{equation}

The total DSF is then expressed as~\cite{siegfried_review}
\begin{equation}
    S_{ee}^\text{tot} (k,\omega) = W_R (k) \delta(\omega) + S_\text{inel}(k, \omega)\,,
\end{equation}
by separating inelastic and quasi-elastic ionic scattering contributions.
In XRTS, the DSF is formally a function of the full momentum vector shift. 
However, for isotropic systems, the DSF only depends on the magnitude $k=|\kv|$, therefore we drop the vectorized form going forward.

The elastic feature, commonly referred to as the Rayleigh weight, for a multi-component system is defined as~\cite{Wuensch_2008}
\begin{equation}
    W_R(k) = \sum_{a,b} \sqrt{x_a x_b} [f_a(k) + q_a(k)] [f_b(k) + q_b(k)] S_{ab}(k)
    \,,
\end{equation}
for the electronic screening cloud $q_j$~\cite{Gericke_2010} and the ionic form factor $f_j$~\cite{Pauling_1932} of species $j=a,b$, with the static structure factors $S_{ab}$ describing ionic correlations between the different species.

Using the Chihara decomposition~\cite{Chihara_1987}, inelastic scattering contributions can then further be split into photons scattering off the free and bound electrons
\begin{equation}
    S_{\text{inel}}(k,\omega) = S^{\text{ff}}(k,\omega) + S^{\text{bf}}(k,\omega) \,.
\end{equation}

The free-free contributions (ff) are typically modelled using simple models for the uniform electron gas~\cite{mermin_prb_70} multiplied by the free charge state $Z^f$ as
\begin{equation}
    S^{\text{ff}}(k,\omega) = Z_f S_{ee}^0 (k,\omega) \,,
\end{equation}
in terms of the DSF of the uniform electron gas $S_{ee}^0$.
One of the most commonly used models to estimate the free-free dielectric function, which is used to calculate the DSF using the dissipation-fluctuation theorem, is the random phase approximation~\cite{pines}.
Local field corrections are then added to account for electronic correlation effects~\cite{Hubbard_1957,Fortmann_2010}.
Electron-ion collisions can be included in the dielectric response using the Mermin function~\cite{mermin_prb_70}, where a dynamic collision frequency is commonly estimated using the Born approximation(see e.g.~\cite{reinholz2000}).

Similarly, the bound-free feature can be described as
\begin{equation}
     S^{\text{bf}}(k,\omega) = \sum_a Z_a^b S_{ea}^{\text{core}}(k,\omega) \,,
\end{equation}
in terms of the bound charge state $Z_b = \text{AN} - Z_f$ and the contributions of each occupied sub-shell $S_{ea}^\text{core}$.
Typically, the bound-free contribution is modelled using the well-known impulse approximation~\cite{Schumacher_1975,siegfried_review}, with continuum lowering effects by surrounding bound electrons accounted for using a simplified ionization potential depression (IPD) model.
The inverse process free-bound is accounted for using the detailed balance relation~\cite{boehme2023evidence}:
\begin{equation}
    S_{ee}^\text{fb}(k,\omega) = S_{ee}^\text{bf}(k,\omega) \exp{\left(-\frac{ \omega}{k_B T} \right)} \,.
\end{equation}

\section{Method}\label{sec:method}

Even though the theoretical description of the DSF in the Chihara decomposition is well-known and widely published (see e.g.~\cite{siegfried_review,Gregori_PRE_2003,Gregori_2007}), there is a distinct lack of openly available codes, making it difficult to iterate and improve upon the method. 

Here, we introduce a newly written Python code for ``x-ray diagnostics, analysis, verification and exploration`` (xDAVE) which includes commonly used models to describe the individual scattering components as lined out above.
The code is open-source and available on GitHub~\cite{xdave_repo}.
\texttt{xDAVE} is written entirely in Python 3, using only publicly available Python packages that are consistently maintained by a large community, such as \texttt{NumPy}~\cite{harris2020array} and \texttt{SciPy}~\cite{scipy_api}.
This was done to avoid potential version conflicts, but also to improve portability and ensure the code can be run on a variety of systems without a complicated installation process.
The individual components describing elastic, bound-free and free-free scattering processes are implemented using a class-based structure, with the physical quantities being tracked in a separate class called the \texttt{PlasmaState}, which also consistently calculates attributes like the Fermi wavenumber or the thermal wavelength.
\texttt{xDAVE} only requires inputs for the electron and ion temperature, charge state and mass density, with optional inputs given for effects like the IPD or the binding energies of each shell.

To account for mixed species, contributions like the bound-free feature are calculated for each state separately~\cite{Wuensch_thesis,Wuensch_2008}. 
For example, given a mixture of $50$ \%C and $50$ \% H, with charge states $Z_C=3.5$ and $Z_H=0.5$, four states are calculated for $H^0$, $H^{1+}$, $C^{3+}$ and $C^{4+}$ with their relative contributions accounted for using partial densities.
Similarly, static structure factor calculations using the hypernetted chain (HNC) approximation~\cite{hansen2013theory} are done to account for correlations between all four ion species in this particular example.
A mean \texttt{PlasmaState} is calculated for the mixed material, for instance to keep track of the total number of free electrons independent of ion species, which is used to calculate the free-free contribution to the DSF assuming a uniform electron gas.

For more details on the implemented models and the code structure, please refer to the supplementary material.
The code is available as a GitHub repository~\cite{xdave_repo}.

XRTS experiments are typically analysed by applying a forward model to estimate parameters like temperature and density from a best-fit optimization.
One of the largest uncertainties in any XRTS analysis is the shape of the IF, and it has a notable effect on inferred conditions; e.g. temperature from detailed balance. Using a ray-tracing code to simulate the crystal response instead of applying an analytic form reduces uncontrolled assumptions in the IF and provides a more realistic scattering spectrum.
The ray-tracing code \texttt{HEART}~\cite{Gawne_2025_heart} was chosen to for this since it is reasonably fast and can be easily parallelized to fit into a typical optimization framework.
Further, since both codes are written in Python 3, their coupling is straightforward.
The output from \texttt{xDAVE} is the DSF convolved with the source profile, which serves as the input for \texttt{HEART} to then ray-trace a detector image.
Similarly to analysing an experimental spectrum, a lineout is then obtained by defining a region of interest and integrating over the signal.
Uncertainties in the detector calibration are removed by directly calibrating the energy axis to the detector pixels.
\texttt{HEART} calculates the crystal response for each photon energy, including the absorption and scattering cross-sections, and the rocking curve. 
Further, as a full 3D ray-tracer, \texttt{HEART} intrinsically includes geometric effects, thus removing uncertainties due to the solid angle coverage of the spectrometer and the relative source-crystal-detector positions.

To obtain best-fit parameters through an optimization process of fitting the synthetic spectrum to experimental data, we have applied the Nelder-Mead algorithm using the Python package \texttt{lmfit}~\cite{lmfit_version1-3-4}, with a loss function defined as
\begin{equation}
    p_\text{loss} = (P_\text{exp} - P_\text{xdave})^2 \,.
\end{equation}
Error bars are estimated using the MCMC package \texttt{emcee}~\cite{emcee_package}.
The log-likelihood function here is defined as
\begin{equation}
    p = -\frac{1}{2} \left[ \frac{(P_\text{exp} - P_\text{xdave})^2 }{\sigma^2} + \log(\sigma^2) \right] \,,
\end{equation}
where $\sigma^2 = \Delta P_\text{exp}^2 + P_\text{xdave}^2 \exp(2 \log f)$, $\log f$ is a free parameter and the experimental error $\Delta P_\text{exp}=1\times10^{-4}$ is kept constant.

\texttt{xDAVE} is run in its default mode, where the Lindhard dielectric function calculation uses an effective static local field correction~\cite{Dornheim_PRB_ESA_2021,Dornheim_PRL_2020_ESA} to model the free-free component; the bound-free and free-bound are modelled using the Impulse Approximation~\cite{Schumacher_1975} and detailed balance respectively~\cite{boehme2023evidence}.
To account for continuum lowering effects, we use the Crowley IPD model~\cite{Crowley_2014_ipd}.
The static structure factors for the Rayleigh weight calculations are calculated using a HNC solver~\cite{Wuensch_2008} with the Yukawa potential to model the electron-ion and ion-ion pseudo-potentials.
Screening of the core electrons by the free ones is taken into account using the finite wavelength screening model~\cite{Chapman_2015} and atomic form factor are obtained using the Pauling-Sherman approximation~\cite{Pauling_1932}.
These are the default options in the code, a list of the full modelling capabilities available is given in the supplementary material.

\section{Results}\label{sec:results}

In this section, we first validate \texttt{xDAVE} by re-analysing an XRTS spectrum obtained from isochorically heated beryllium at the OMEGA Laser Facility~\cite{Doeppner_2009}, before describing predictions for potential experiments at an XFEL facility and the National Ignition Facility~\cite{Kraus_2016}.

\subsection{OMEGA Beryllium foil experiment}\label{subsec:be-omega}

To demonstrate the capability of our new code as a forward model, we now turn to analysing a previously published OMEGA XRTS spectrum of a isochorically heated beryllium foil~\cite{Doeppner_2009}.
In this work, the focus is on reproducing the analysis, rather than introducing new physics.
Initially, the determined conditions were an electron temperature of $18 \pm 4 \, \unit{eV}$, an ion temperature of $6 \, \unit{eV}$ and a mass density of $\rho=1.2 \pm 0.2 \, \unit{g/cc}$.
An estimated ionization degree of $Z=2.3$ was extracted in comparison against previous analysis~\cite{Glenzer_PRL_2007} and radiation-hydrodynamics simulations~\cite{Doeppner_2009}.
Sch\"orner~\emph{et al.}~\cite{Schoerner_PRE_2023} subsequently analysed the same dataset using TD-DFT and found similar temperatures but a slightly higher density at $1.8 \unit{g/cc}$ and a lower ionization degree at $Z=2.14$.
Model-free temperature analysis using the ITCF further found an electron temperature of $T_e=19.0 \pm 1.5 \, \unit{eV}$~\cite{Schoerner_PRE_2023}.

Here, we extend the original Chihara analysis to simultaneously fit all parameters, including obtaining an estimate for the charge state using improved models of the static structure factor.
The models for the Rayleigh weight implemented here have been shown to better match advanced simulations like DFT or MD~\cite{Wuensch_2008,Johnson_2024}, which allows for a better estimate of the charge state than using the original one-component model~\cite{Gregori_2007}.
\texttt{xDAVE} has a dedicated HNC solver package to estimate the static structure factors, which has been shown to capture correlation effects in the warm dense matter regime more accurately than the one-component like case~\cite{Wuensch_2008}.
Additionally, more advanced screening models like the finite wavelength screening~\cite{Gericke_2010} have also been shown to play a significant role in the warm dense matter regime~\cite{Chapman_2015}.
Both play a substantial role in the determination of the Rayleigh weight and thus the strength of the elastic peak.

\begin{figure}
    \centering
    \includegraphics[width=0.99\linewidth]{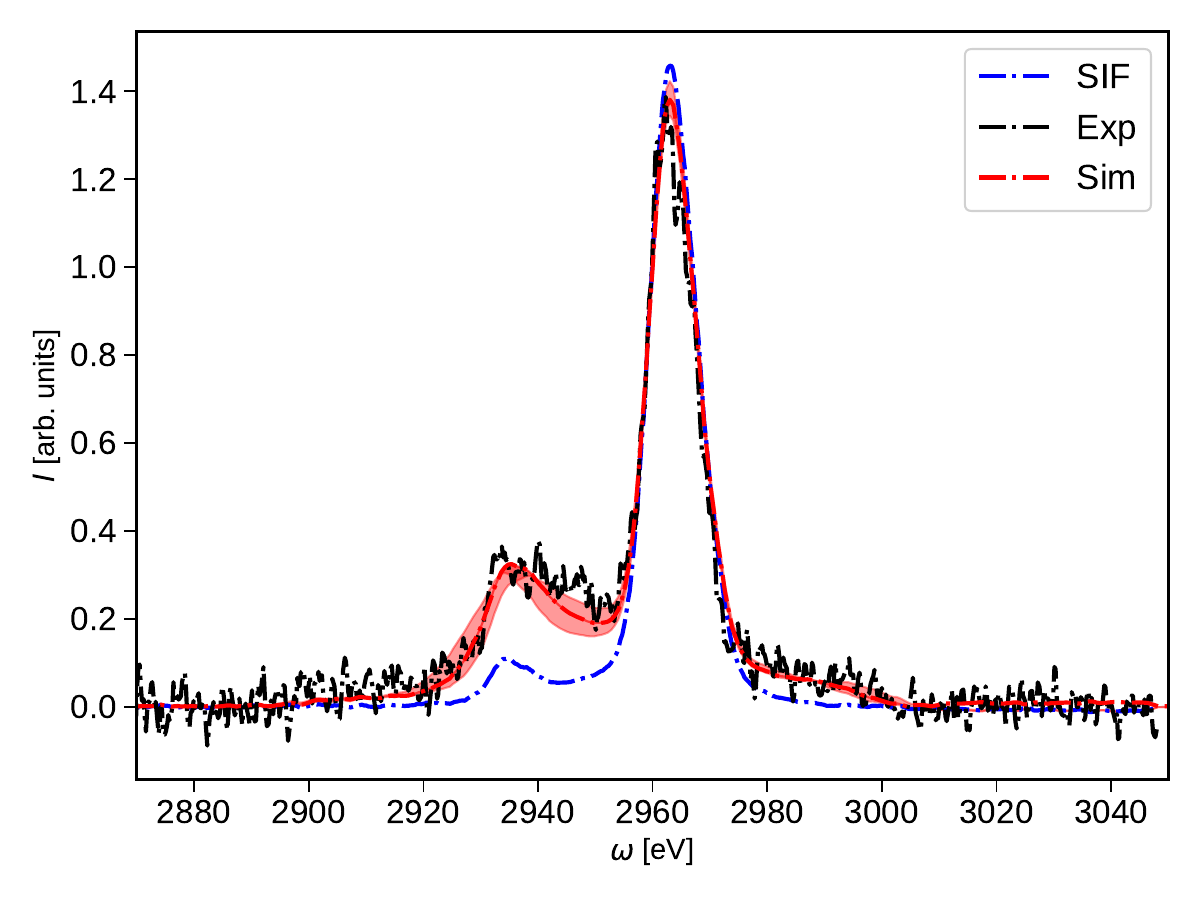}
    \caption{Optimized fit for a Beryllium XRTS measurements at OMEGA. Best fit parameters found are $T_e=16.9 ^{+4.7} _{-0.9} \, \unit{eV}$, $T_i = 7.3 ^{+2.3} _{-1.1} \, \unit{eV}$, $\rho=1.9 ^{+0.4} _{-0.2} \, \unit{g/cc}$ and $Z=2.4 ^{+0.1} _{-0.4}$. Error bars obtained from MCMC are indicated in the shaded red area.
    }
    \label{fig:be-omega-fit}
\end{figure}

Fig.~\ref{fig:be-omega-fit} shows the results of an optimization run to find the best fit parameters for the spectrum, with the best-fit parameters found as $T_e=16.9 ^{+4.7} _{-0.9} \, \unit{eV}$, $T_i = 7.3 ^{+2.3} _{-1.1} \, \unit{eV}$, $\rho=1.9 ^{+0.4} _{-0.2} \, \unit{g/cc}$.
For consistency with the original approach, both spectra are normalised to the elastic peak at the probe energy of $E_b=2.96 \unit{keV}$.
The slightly smaller electron temperature found here is likely due to the inclusion of the free-bound component in our analysis, but within the error bars is still in agreement with the model-free temperature found by Sch\"orner~\emph{et al.}~\cite{Schoerner_PRE_2023}.
Previous work has shown that this can reduce the estimated temperature by correctly modelling the up-shifted regime using detailed balance~\cite{boehme2023evidence}.
The extracted ionization degree is in very good agreement with the original estimate obtained from the analysis of similar experiments and radiation-hydrodynamic simulations~\cite{Doeppner_2009}, while being slightly higher than the value obtained by Sch\"orner~\emph{et al.}~\cite{Schoerner_PRE_2023} using the Thomas-Reiche-Kuhn sum rule.

The error bars estimated here using MCMC are shown in the shaded area in Fig.~\ref{fig:be-omega-fit}.
Particularly for the temperatures, the estimated error is comparably large due to a combination of the spectral noise and the stability of fitting against four parameters simultaneously~\cite{Kasim_POP_2019}.

These results show that using \texttt{xDAVE}, we can reproduce previously estimated parameters, while additionally obtaining estimates for the charge state using a Chihara forward model.
Further, using the advanced models in \texttt{xDAVE} for the static structure factor and screening cloud calculations, we can obtain density estimates that more closely match the \emph{ab initio} TD-DFT results~\cite{Schoerner_PRE_2023} and the solid density of beryllium that is expected for an isochoric heating experiment.

\subsection{Predicting XRTS experiments at EuXFEL}\label{subsec:eu-xfel}

Combining a fast XRTS code with a ray-tracing code can give insights into the planning and designing of future experiments at a variety of experimental facilities.
To demonstrate, we consider a possible plastic scattering spectra for a typical XRTS experiment at an XFEL facility like the European XFEL~\cite{Zastrau_2021}.
For a beam energy of $8.125 \unit{keV}$ we consider the standard von Hamos setup with a HAPG crystal and Jungfrau detector~\cite{Preston_2020} for multiple scattering angles.

\begin{figure}
    \centering
    \includegraphics[width=0.99\linewidth]{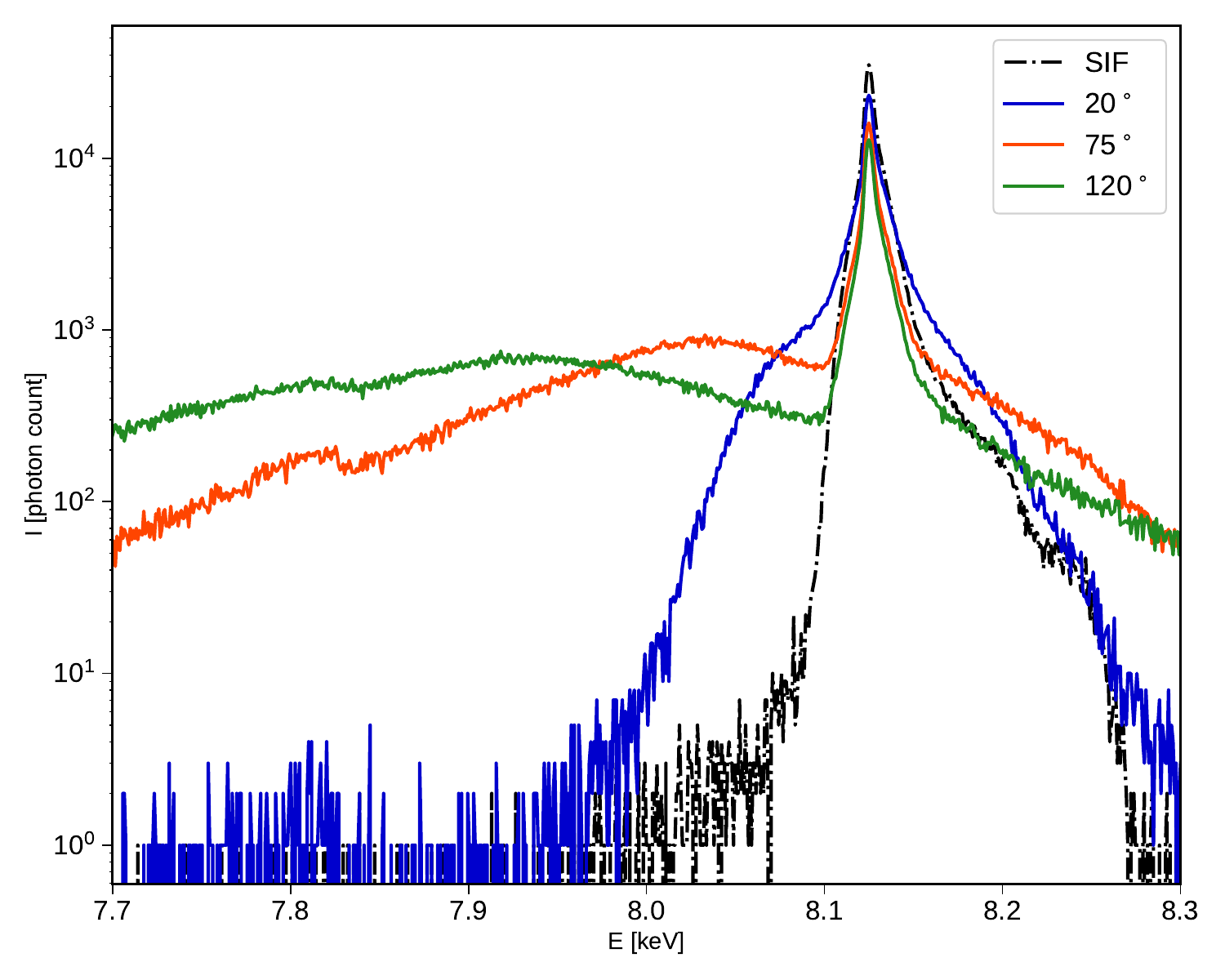}
    \caption{Predicted CH scattering spectrum using a typical XFEL seeded beam as the source (black dashed) compared at three different scattering angles for $T=70 \unit{eV}$, $\rho=1.9 \unit{g/cc}$ and a carbon charge state of $Z_C=3.1$.
    The material is comprised of $20\%$ fully-ionized Hydrogen giving a mean charge state of $2.68$.
    The ray-traced SIF is plotted in black.
    }
    \label{fig:ch-heart-spectrum-multiple-angles}
\end{figure}

Previous experiments have demonstrated densities of several times the solid density~\cite{Kraus_2025} and temperatures of several tens of eVs~\cite{Martin_2025,boehme_2025_preprint}.
While the source spectrum is regularly recorded, the instrument function is challenging to accurately measure~\cite{Gawne_2024-09}, leaving uncertainties particularly in the upshifted part of the spectrum where asymmetric wings of a typical HAPG instrument function can lead to overestimated temperatures when not properly accounted for~\cite{Gawne_2024-09}.
Fig.~\ref{fig:ch-heart-spectrum-multiple-angles} gives an example of a spectra at multiple scattering angles produced using the described  \texttt{xDAVE} + \texttt{HEART} setup.
Here, we use an analytic source simulated to match a seeded beam profile that can realistically be achieved at the European XFEL~\cite{Zastrau_2021}.
This was achieved by combining a wide Gaussian distribution with full width at half maximum (FWHM) of $20 \unit{eV}$ as the pedestal and a narrow Gaussian with FWHM=$1\unit{eV}$ as the seed.
The relative intensity of the pedestal is set at $5\%$ compared to the seed.

The three angles shown in Fig~\ref{fig:ch-heart-spectrum-multiple-angles} correspond to the collective, intermediate and non-collective scattering regime.
In the collective regime, the electronic response is dominated by correlated motion rather than single-particle effects.
The probe wavelength is long compared to the screening length, as opposed to the non-collective, where the probing wavelength is short and electrons effectively scatter as single particles.
For the collective case (blue) at $20^\circ$, we see the characteristic plasmon downshifted from the elastic peak.
The non-collective scattering spectrum at $120 ^\circ$ shows the typical wide Compton peak.
We can also observe the impact of ionization potential depression on the relative position of the carbon $L$-shell at $\omega \approx 7.81 \unit{keV}$.
The asymmetry of the instrument function is clearly visible in the upshifted energy regime for all scattering angles.
Since electron temperature estimates are commonly derived from detailed balance, which relates the up- and downshifted intensity, this asymmetry of the SIF has a large effect on the extracted values if it is not properly accounted for.

By deconvolving the spectra with a ray-traced SIF using the model-free ITCF method, we can estimate conditions required to extract accurate temperatures directly from a spectrum~\cite{Dornheim_T_2022}.
This is achieved by using the well-known deconvolution theorem in imaginary time $\tau$~\cite{Dornheim_T_follow_up}
\begin{equation}
    F(k,\tau) = \cfrac{ \mathcal{L}[S(k,\omega) * R(\omega)]}{\mathcal{L}[R(\omega)]} \,,
\end{equation}
where the ITCF is defined as the double-sided Laplace transform of the DSF
\begin{align}\label{eq:laplace-transform}
    F(k,\tau) &= \mathcal{L}[S_{ee}(k,\omega)]  \\ \nonumber
    &= \int_{-\infty}^{\infty} d\omega S_{ee}(k,\omega) \exp{(-\tau \omega)} \,.
\end{align}
The temperature is then estimated using the detailed balance relation which is expressed as a symmetry relation around the inverse temperate $\beta$ in imaginary time
\begin{equation}
    F(k,\tau) = F(k,\beta - \tau) \,.
\end{equation}
In practice, because of the limited detector range, the integral in Eq.~\ref{eq:laplace-transform} is truncated with respect to the integration limit $x$ and the temperature is estimated from the converged limit
\begin{equation}
    \lim_{x \rightarrow \infty} F_x(k,\tau) = F(k,\tau) \,.
\end{equation}

\begin{figure}
    \centering
    \includegraphics[width=0.95\linewidth]{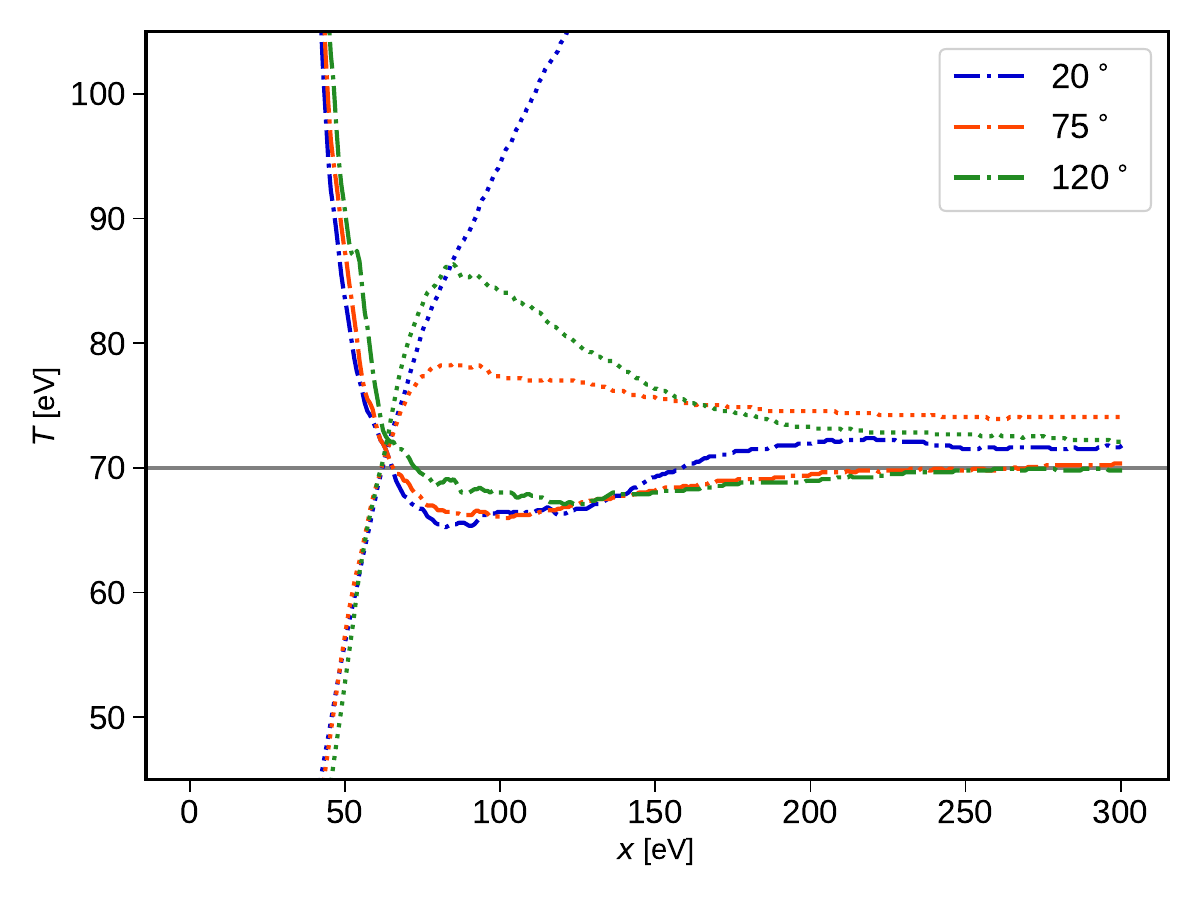}
    \caption{Model-free temperature analysis for the CH spectra plotted in Fig.~\ref{fig:ch-heart-spectrum-multiple-angles}. The Laplace transform of the simulated spectrum (dotted) is shown in comparison against the deconvolved ITCF (dashed-dotted) for three different scattering angles for increasing integration limit $x$.
    The expected temperature of $70 \unit{eV}$ is shown in the gray solid line.
    }
    \label{fig:itcf-temperature-analysis}
\end{figure}

The analysis is shown in Fig.~\ref{fig:itcf-temperature-analysis}, with the input temperature of $T=70 \unit{eV}$ indicated by the gray solid line.
As the collective scattering signal decays much quicker in the upshifted regime than the non-collective, extracting a model-free temperature is much more challenging.
For the $75^\circ$ and $120^\circ$ scattering angles, we see a clear convergence to the correct temperature for the integration limit $x > 200 \unit{eV}$.
This allows detailed considerations in the design of future experiments, including estimating the required number of shots to obtain a highly resolved spectrum by taking into account the photon number, and correctly predicting the shape of the SIF to reduce uncertainty in the temperature estimates.
It can also inform future experiments on the required dynamic range to accurately resolve temperature measurements.

\begin{figure}
    \centering
    \includegraphics[width=0.49\linewidth]{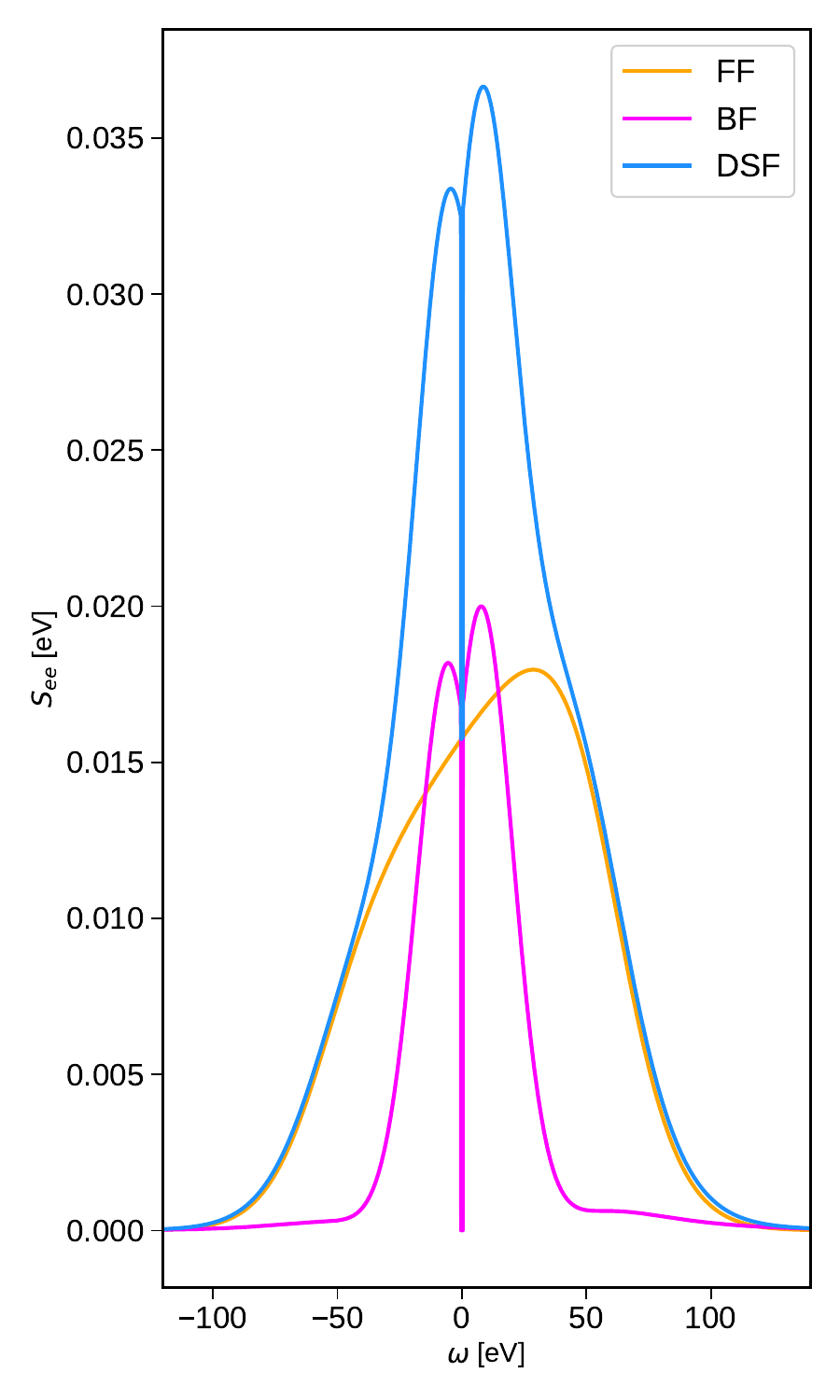}
    \includegraphics[width=0.49\linewidth]{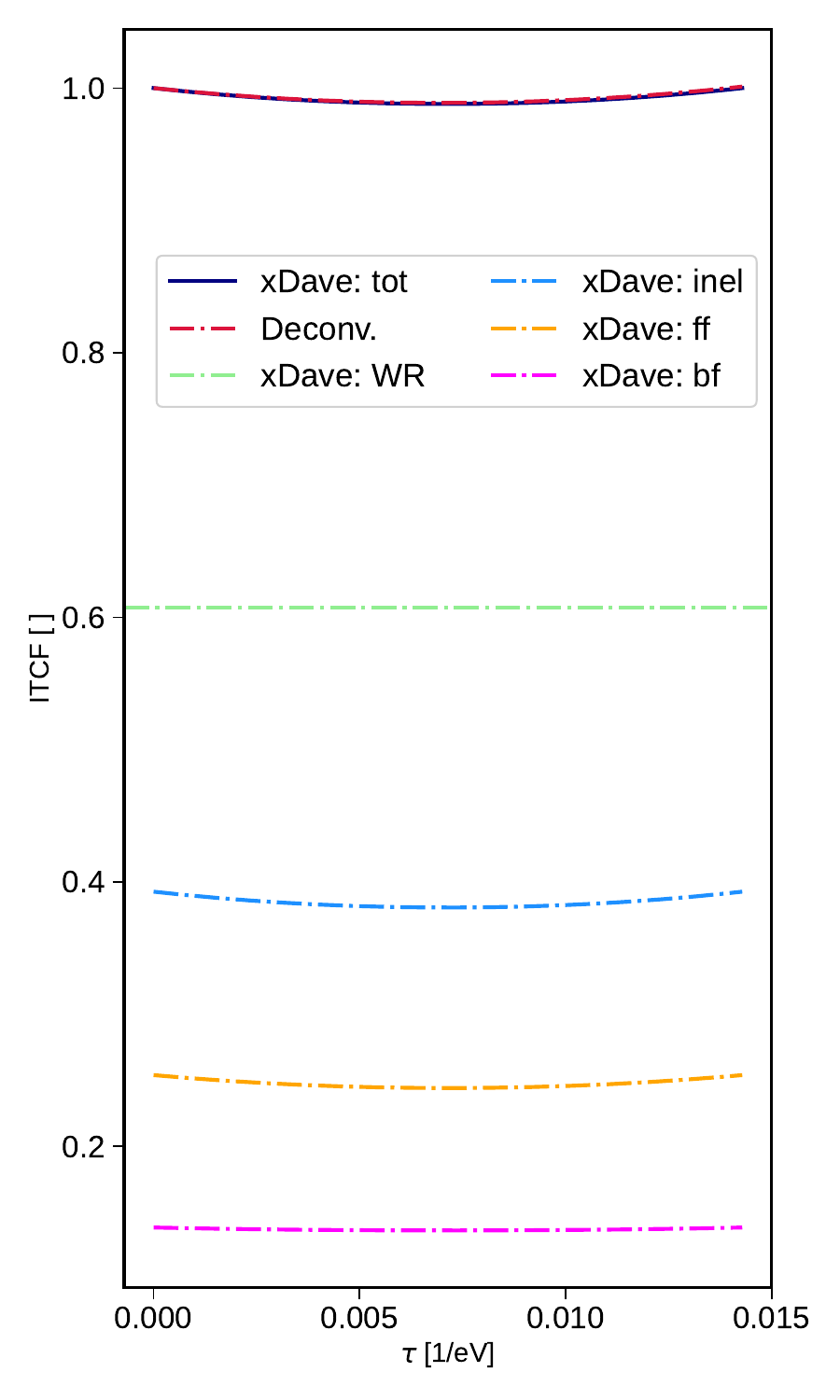}
    \caption{\texttt{xDAVE} result for the dynamic structure factor (left) and the corresponding imaginary-time correlation function (right) for the conditions described in Fig.~\ref{fig:ch-heart-spectrum-multiple-angles} for a scattering angle of $20^\circ$.
    We show the individual components to the inelastic feature in the dynamic structure: free-free (orange), bound-free (pink) and total (blue).
    The ITCF plot also shows the Rayleigh weight as a constant in $\tau$-space (purple).
    The deconvolved ITCF from the \texttt{HEART} spectrum is shown in red at an integration limit of $x=120 \unit{eV}$.
    }
    \label{fig:dsf-xdave-ch-20deg}
\end{figure}

\begin{figure}
    \centering
    \includegraphics[width=0.49\linewidth]{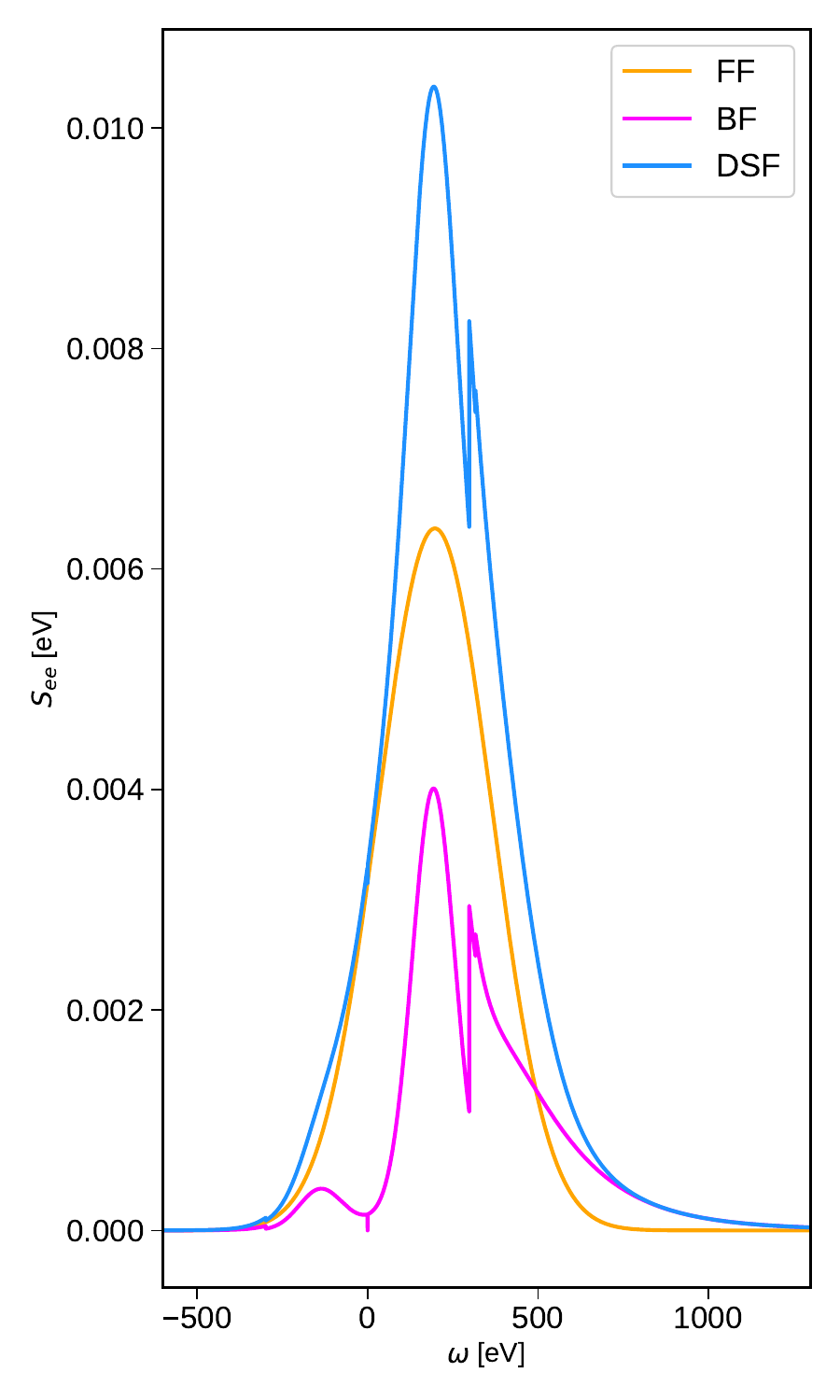}
    \includegraphics[width=0.49\linewidth]{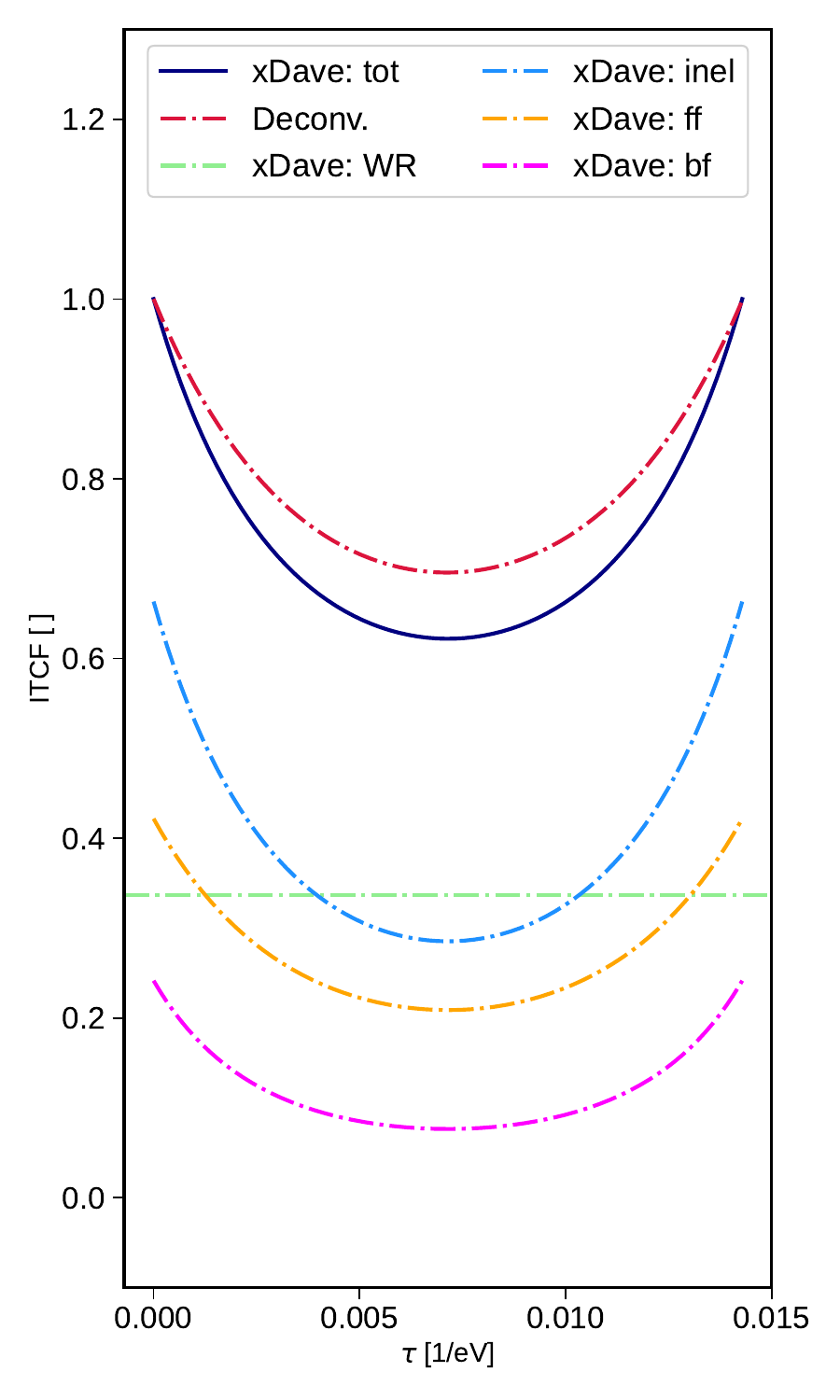}
    \caption{\texttt{xDAVE} result for the dynamic structure factor (left) and the corresponding imaginary-time correlation function (right) for the conditions described in Fig.~\ref{fig:ch-heart-spectrum-multiple-angles} for a scattering angle of $120^\circ$.
    We show the individual components to the inelastic feature in the dynamic structure: free-free (orange), bound-free (pink) and total (blue).
    The ITCF plot also shows the Rayleigh weight as a constant in $\tau$-space (purple).
    The deconvolved ITCF from the \texttt{HEART} spectrum is shown in red at an integration limit of $x=600 \unit{eV}$.
    }
    \label{fig:dsf-xdave-ch-120deg}
\end{figure}

Details of the corresponding \texttt{xDAVE} simulations are given by Fig.~\ref{fig:dsf-xdave-ch-20deg} for the collective and Fig.~\ref{fig:dsf-xdave-ch-120deg} for the non-collective case.
The separate bound-free and free-free scattering contributions to the DSF are given by the respective left plot, whereas the right shows corresponding ITCFs.
The Rayleigh weight is also plotted as a constant (green) line in the imaginary-time domain for each case.
For both scattering angles, the IPD was estimated using Crowley's model~\cite{Crowley_2014_ipd} for the $C^{3+}$ as $\Delta_\text{IPD} = -63.9 \unit{eV}$ and $\Delta_\text{IPD}=-75.4 \unit{eV}$ for the $C^{4+}$ shell.

The collective scattering case at $20^\circ$ (Fig.~\ref{fig:dsf-xdave-ch-20deg}) shows the narrow spectral width of the spectrum and the plasmon characteristic for such small scattering wavevectors.
Having access to the ITCF from \texttt{xDAVE} allows a direct comparison against the experimentally obtained one.
We observe very good agreement between the deconvolved ITCF and the one obtained from \texttt{xDAVE} for an integration limit of $x=120 \unit{eV}$.
In contrast, for the non-collective case at $120^\circ$, the deconvolved ITCF and the \texttt{xDAVE} one disagree significantly for an increased integration limit of $x=600 \unit{eV}$.
This is likely due to the limited available detector range as seen in Fig.~\ref{fig:ch-heart-spectrum-multiple-angles}.
For the non-collective case, we can observe that the spectrum has not decayed sufficiently to capture the full ITCF in the downshifted energy range.
Additionally, the SIF is no longer well-defined at these energy ranges, introducing noise to the deconvolution.
Therefore, even for a large integration limit of $x=600 \unit{eV}$, we cannot fully resolve the whole spectrum preventing a deconvolution due to the broad nature of the inelastic Compton peak.
This would affect estimates like the normalization using the f-sum rule~\cite{Dornheim_SciRep_2024} or the extraction of the Rayleigh weight for density estimates~\cite{Dornheim_2025_POP}.
Nevertheless, the position of the minimum in the ITCF and the temperature analysis (see Fig.~\ref{fig:itcf-temperature-analysis}) indicate that we can sufficiently resolve the upshifted feature to obtain an accurate temperature measurement from detailed balance.
Further, this analysis allows us to carefully plan future experiments, taking into account limited detector range and the impact of the SIF on the spectral deconvolution.

\subsection{NIF Beryllium implosions}\label{subsec:nif-implosions}

Lastly, the same ray-traced analysis is applied to the implosion platform at the NIF~\cite{Kraus_2016}, where XRTS measurements have been performed to extract time-resolved measurements of compressed Beryllium~\cite{Tilo_Nature_2023}.
The scattering spectra are collected on the Mono-Angle Crystal (MACS) spectrometer~\cite{Doeppner_RSI_2014}. 
This consists of a gated, four-strip multi-channel plate (MCP) detector, which is typically used to measure spectra before and after the time of peak x-ray emission; along with a cylindrically-bent HOPG crystals used to disperse the x-rays in a Hall-like geometry~\cite{Hall_Geometry}.
The capsule is compressed using 184 of NIF's 192 laser beams, with the remaining eight utilized to generate $9 \unit{keV}$ x-rays from zinc He-$\alpha$ emissions from a foil.

\begin{figure}
    \centering
    \includegraphics[width=0.99\linewidth]{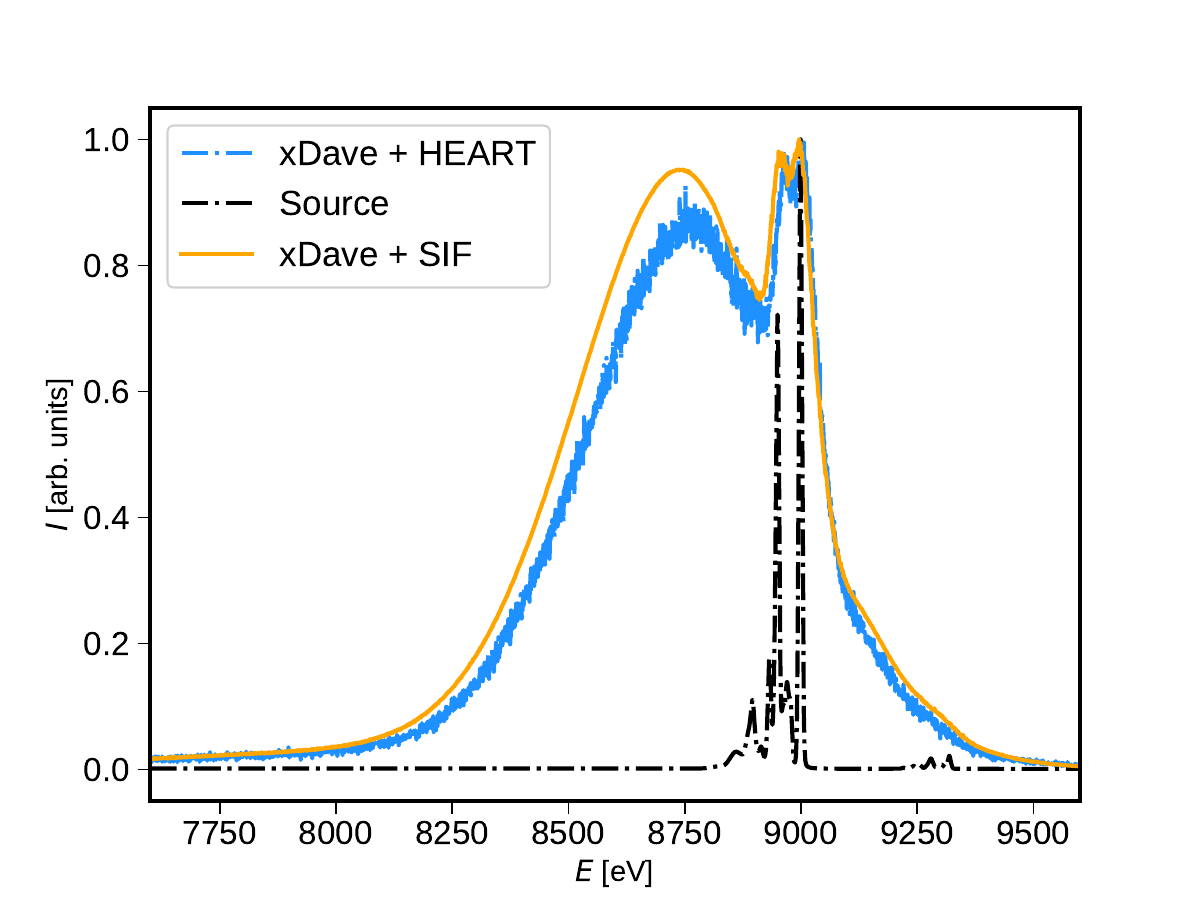}
    \includegraphics[width=0.99\linewidth]{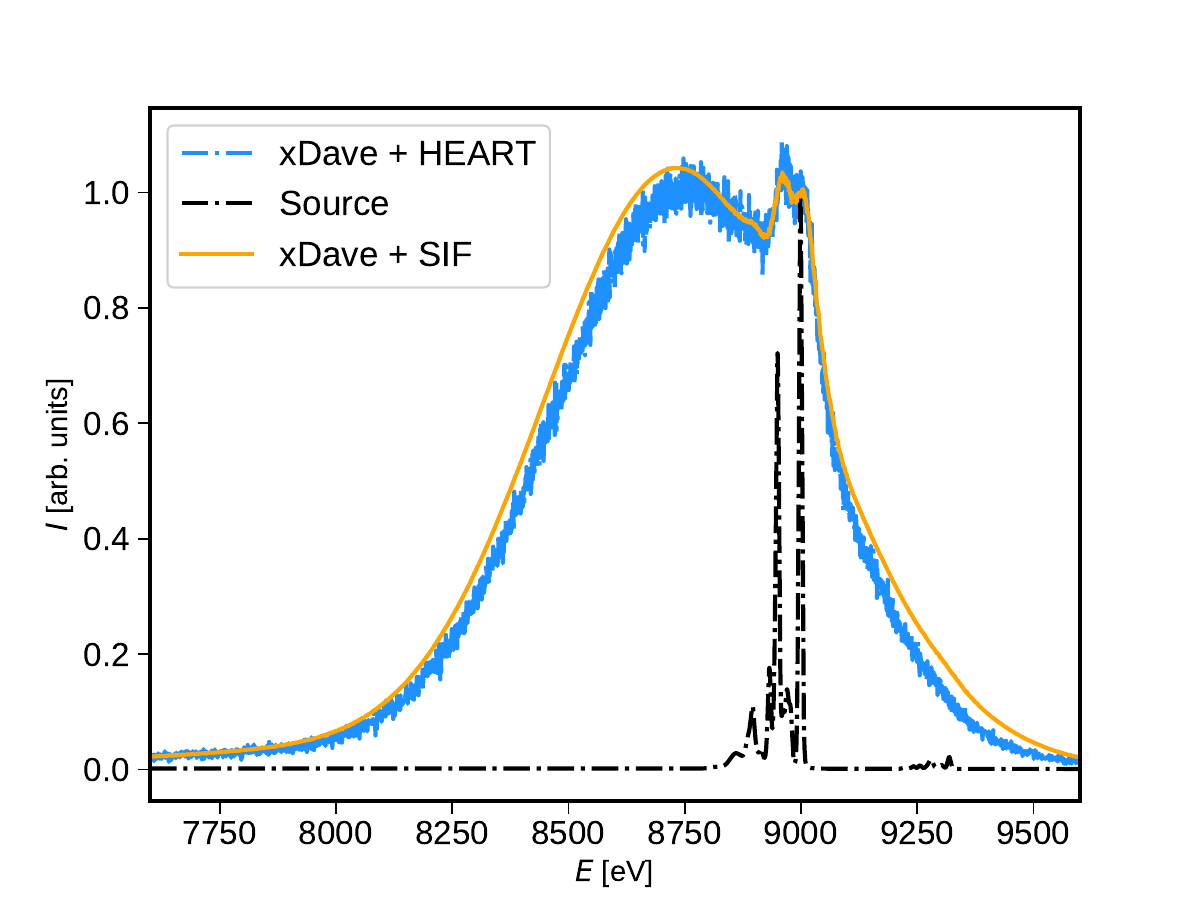}
    \caption{XRTS spectrum for a NIF Be implosion experiment.
    Top: Spectrum estimated before time of peak x-ray emissions, conditions set as $T=110 \unit{eV}$, $\rho=6. \unit{g/cc}$ and $Z=3.1$. \\
    Bottom: Spectrum estimated after time of peak x-ray emissions, conditions set as $T=160 \unit{eV}$, $\rho=20. \unit{g/cc}$ and $Z=3.8$. \\
    The source profile obtained from FLYCHK is shown in black.
    Spectra are normalised to the elastic peak at $9 \unit{keV}$.
    }
    \label{fig:nif-be-results}
\end{figure}

The zinc emission lines are modelled using FLYCHK~\cite{Chung_2005_flychk} with a temperature of $4 \, \unit{keV}$ and an electron density of $1 \times 10^{21} \, \unit{g/cc}$~\cite{MacDonald_POP_2022}, and are shown in black for both cases in Fig.~\ref{fig:nif-be-results}.
The conditions for both spectra here were chosen to present parameters found by D\"oppner ~\emph{et al.}~\cite{Tilo_Nature_2023} and subsequent analysis to constrain the density estimates by Dornheim~\emph{et al.}~\cite{Dornheim_2025_POP,Dornheim_2025_NatCommun} and others~\cite{Dharma_wardana_2025,Zeng_2025}.
The first strip is assumed to have $T=110 \, \unit{eV}$, $\rho=6 \, \unit{g/cc}$ and $Z=3.1$ and is given by the top panel of Fig.~\ref{fig:nif-be-results}.
The second shot is taken to correspond to spectra measured after peak x-ray emission time is assumed to have plasma parameters $T=160 \, \unit{eV}$, $\rho=20 \, \unit{g/cc}$ and $Z=3.8$.

For both cases, we see significant deviations between the ray-traced spectrum and the convolution with a SIF. In particular, the position of the inelastic peak and the ratio of elastic-to-inelastic scattering appear notably different depending on whether the spectrum is ray-traced or a SIF is used.
Even though the SIF used here was created by ray-tracing the source spectrum using \texttt{HEART}, the geometric layout of the spectrometer significantly impacts both the width and the intensity of the IF across the spectral range, which is then observed in the detected spectrum.
In addition, the ray tracing approach removes uncertainties due to the calibration of a spectrometer, particularly for a complicated setup like the NIF geometry.
Due to the use of a cylindrical crystal in a Hall geometry rather than in von H\'amos geometry~\cite{vonHamos_Geometry} (owing to space constrains in the vacuum chamber at the NIF), the spectrometer dispersion is much more complicated along the detector and difficult to extract. 
This introduces uncertainties in the calibration of the energy axis, and therefore the actual energy shift of features in the scattering spectrum.
Instead of relying on analytic dispersion relations or fits using the elastic peak, the spectrometer can be directly calibrated using \texttt{HEART}, or by comparing simulated spectra pixel-to-pixel with the experimental spectrum.
Effects like filter transmission can also be accounted for within the ray-tracing approach, as \texttt{HEART} allows the user to add filters ahead of the detector and crystal. 
The loss of photons due to their actual angle of incidence on a filter is thereby accounted for~\cite{Gawne_2025_heart}.

These effects can impact conditions derived from a forward-fit and should be accounted for in the analysis of future experiments.
Particularly the difference in the upshifted regime can lead to an overestimation of the temperature using a detailed balance analysis.
Overestimating the width of the Compton feature as seen for the convolved SIF case can also lead to an overevaluation of the free electron number density -- and therefore the charge state and mass density -- since the inelastic feature becomes increasingly broadened for a larger number of free electrons~\cite{siegfried_review}.
We therefore advocate for future experiments to include ray-tracing in their analysis framework.
This would allow a direct comparison of the synthetic detector image obtained from \texttt{HEART} and \texttt{xDAVE} against experimental results, removing error introduced in the lineout analysis due to corrections for the solid angle or the dispersion relation.
It would further allow the analysis of experimental spectra where defocusing effects complicate the forward-fit modelling, or allow for the self-consistent treatment of source broadening for large targets.
Additional geometric effects such as the range of scattering vectors covered by the crystal can also be readily integrated into a ray-tracing framework.
\texttt{HEART} natively parallelizes ray-tracing simulations over multithreading and \texttt{MPI}~\cite{Gawne_2025_heart}, making it a viable inclusion in the large-scale parameter scans that are required to extract best-fit parameters in an optimization, therefore enabling an improved forward-fitting approach.

Future work will focus on quantifying the difference between the ray-traced and analytic approach for different points in parameter space.
As temperature and density significantly impact the difference between both methods, as seen in Fig.~\ref{fig:nif-be-results}, a detailed analysis can give further insights into systematically estimating the divergence depending on plasma conditions.

\section{Summary and Outlook}

In this work we have presented a new XRTS code relying on the Chihara decomposition to quickly estimate dynamic structure factors for the warm dense matter regime.
The code presented here is openly available on GitHub~\cite{xdave_repo}.
The current version of \texttt{xDAVE} includes most commonly used chemical models.
A full overview of the models contained in \texttt{xDAVE} (along with their validations) is given in the supplementary material.

\texttt{xDAVE} has been validated by forward-fitting to a previously analysed isochorically heated Beryllium spectrum at the OMEGA Laser Facility~\cite{Doeppner_2009}.
Our analysis inferred similar conditions to the original, but the estimated mass density more closely matches recent DFT simulations~\cite{Schoerner_PRE_2023}, which we attribute to the improved modelling capabilities for the elastic peak used here.

We have further established that ray-tracing DSFs obtained from \texttt{xDAVE} can give insights into the planning of future experiments at both XFEL facilities and the National Ignition Facility.
Potential scattering spectra for a typical XFEL experiment of laser-heated CH were shown to demonstrate the asymmetry of the instrument function and the impact of the scattering angle on the temperature estimate using the model-free ITCF analysis.
This framework can reduce uncertainties in the planning of future experiments, by not only taking into account physical limitations like the detector range but also estimating the required photon statistics to extract temperatures using detailed balance.

Subsequent application of the combined ray-tracing approach to the more complicated geometry of the NIF's MACS spectrometer further demonstrated the necessity of accounting for effects like the layout and geometry of the spectrometer components, as well as removing uncertainties in the calibration, for the analysis of future implosion experiments.
The difference, particularly in the upshifted regime from which temperature is commonly estimated using the detailed balance relation, could lead to an overestimation in the temperature. This is exacerbated by both solid angle corrections and the asymmetry of the SIF.
In conclusion, we established that the use of a convolved SIF -- even a ray traced one -- shows significant differences when compared to the fully ray-traced spectrum, indicating future forward-fits should be done using ray-tracing in the model pipeline.
Future work will focus on quantifying the differences observed and conditions extracted for different points in the $\rho-T$ parameter space.

The aim of \texttt{xDAVE} is to have a flexible code structure that can be applied to experimental analysis and also serve as a tool in the comparison against ~\emph{ab initio} simulation tools like DFT and PIMC.
Recent work has demonstrated that fitting an XRTS code to PIMC data can provide estimates for both the charge state and ionization potential depression of warm dense hydrogen~\cite{Bellenbaum_2025_itcf}.
This analysis can easily be extended to higher-Z materials or DFT simulations. Having a code like \texttt{xDAVE}, that separately treats the elastic and inelastic features, allows for more flexible analysis.
This also gives a pathway to improving chemical models currently in use, for example to derive more accurate bound-free descriptions by comparing against DFT or average atom modelling.
In addition, because the code allows for quick estimation of dynamic structure factors, it can be readily applied as a default model to obtain DSFs from PIMC simulations via analytic continuation~\cite{JARRELL1996133,Chuna_2025-09,BENEDIXROBLES2026109904}.
Lastly, the Chihara decomposition allows non-equilibrium models~\cite{Gregori_2007} to be used in the modelling, thereby extending the analysis to, for example, isochorically heated experiments that typically contain different electron and ion temperatures~\cite{siegfried_review}.
In combination with the ITCF model-free temperature analysis, which has been shown to be useful in the quantification of non-equilibrium effects~\cite{vorberger2023revealing,Bellenbaum_APL_2025}, this could allow further insights into the modelling of two-temperature XRTS spectra.

Future improvements to the code will include a coupling to \emph{ab initio} data to obtain quantities like the Rayleigh weight from PIMC or DFT~\cite{Dornheim_2025_POP}, or electron-ion collision frequencies to be included in the Mermin dielectric function from DFT-MD~\cite{hentschel_PoP_2023,hentschel_PoP_2025}.
More advanced non-equilibrium modelling capabilities could also include a two-temperature version of the Lindhard dielectric function to model the free-free component~\cite{Vorberger_2018}.
Improvements to the HNC description by, for instance, including bridge functions for the strongly coupled regime can also improve the description of the elastic peak and give more accurate estimates for the one-component plasma.
Additionally, two-temperature multi-component approaches using HNC exist in the SVT approach~\cite{Seuferling_1989}, which could extend the static structure factor calculations to include more complicated analysis

The combined analysis pipeline of ray-tracing DSFs presented here can also be extended to model the full experiment, including particle transport to remove the assumption of analytic source profiles~\cite{Hernandez_preprint_2026}, and the inclusion of hydrodynamics simulations to model the non-homogeneity of conditions found in XRTS experiments~\cite{Kraus_PRE_2016,Poole_POP_2022}.
The modularity of \texttt{xDAVE} makes it an easily extendible code with a variety of applications that can give accessible and fast estimates of DSFs within the chemical picture, and is straightforward to implement in any analysis pipeline.

\section{Supplementary Material}

The supplementary material contains information on the models implemented in the open-source code \texttt{xDAVE} and details on the implementation.
An overview of the different components of the Chihara decomposition and the underlying model assumptions.

\begin{acknowledgements}

\noindent 
HMB would like to thank T. D\"oppner for providing the experimental spectrum and the SIF of the OMEGA beryllium XRTS measurement.
This work was partially supported by the Center for Advanced Systems Understanding (CASUS), financed by Germany’s Federal Ministry of Education and Research and the Saxon state government out of the State budget approved by the Saxon State Parliament. 
This work has received funding from the European Union's Just Transition Fund (JTF) within the project \emph{R\"ontgenlaser-Optimierung der Laserfusion} (ROLF), contract number 5086999001, co-financed by the Saxon state government out of the State budget approved by the Saxon State Parliament. This work has received funding from the European Research Council (ERC) under the European Union’s Horizon 2022 research and innovation programme (Grant agreement No. 101076233, "PREXTREME"). 
Views and opinions expressed are however those of the authors only and do not necessarily reflect those of the European Union or the European Research Council Executive Agency. Neither the European Union nor the granting authority can be held responsible for them. This work has received funding from the German Federal Ministry of Research, Technology and Space (BMFTR) via the ErUM Data project "DEMOS" (05D25CR1).
Tobias Dornheim gratefully acknowledges funding from the Deutsche Forschungsgemeinschaft (DFG) via project DO 2670/1-1.
Computations were performed at the Norddeutscher Verbund f\"ur Hoch- und H\"ochstleistungsrechnen (HLRN) under grant mvp00024.
\end{acknowledgements}


\bibliography{bibliography}
\end{document}